\documentclass[journal,12pt,letterpaper,draftcls,onecolumn]{IEEEtran}%
\usepackage{amsmath}
\usepackage{amsfonts}
\usepackage{cite}
\usepackage{amssymb}
\usepackage{graphicx}
\usepackage[usenames,dvipsnames]{color}
\usepackage{epstopdf}
\usepackage[all]{xy}
\usepackage[]{mcode}

\providecommand{\U}[1]{\protect\rule{.1in}{.1in}}
\pdfoutput=1
\providecommand{\U}[1]{\protect\rule{.1in}{.1in}}

\newcommand{\qed}{\nobreak \ifvmode \relax \else
      \ifdim\lastskip<1.5em \hskip-\lastskip
      \hskip1.5em plus0em minus0.5em \fi \nobreak
      \vrule height0.75em width0.5em depth0.25em\fi}

\begin{document}

\title{\LARGE A Novel Error Performance Analysis Methodology for OFDM-IM}
\author{Shuping Dang, \textit{Member, IEEE}, Guoqing Ma, \textit{Student Member, IEEE}, Basem Shihada, \textit{Senior Member, IEEE}, Mohamed-Slim Alouini, \textit{Fellow, IEEE}
  \thanks{All authors are with Computer, Electrical and Mathematical Science and Engineering Division, King Abdullah University of Science and Technology (KAUST), 
Thuwal 23955-6900, Kingdom of Saudi Arabia (e-mail: \{shuping.dang, guoqing.ma, basem.shihada, slim.alouini\}@kaust.edu.sa).}}

\maketitle

\begin{abstract}
Orthogonal frequency-division multiplexing with index modulation (OFDM-IM) has become a high-profile candidate for the modulation scheme in next generation networks, and many works have been published to analyze it in recent years. Error performance is one of the most important and interesting aspects of OFDM-IM, which has been analyzed in most works. However, most of them employ two key approximations to derive the closed-form expressions of block error rate (BLER) and/or average bit error rate (BER). The first one is to utilize the union bound assisted by the pairwise error probability (PEP) analysis, and the second one is to employ an exponential approximation of Q-function. In this letter, we apply Craig's formula to analyze the error performance for OFDM-IM in place of the exponential approximation. Because Craig's formula is the exact expression of Q-function, the accuracy of analytical results regarding error performance for OFDM-IM can be improved. We examine the proposed error performance analysis methodology based on Craig's formula by investigating both average BLER and BER.
\end{abstract}


\section{Introduction}
\IEEEPARstart{O}{rthogonal} frequency-division multiplexing with index modulation (OFDM-IM) becomes one of the most promising modulation candidates in next generation networks, and has been exhaustively studied in recent years \cite{8004416}. OFDM-IM employs an additional modulation dimension termed the \textit{index} dimension in addition to the amplitude and phase of constellation symbol to convey extra information bits by subcarrier activation patterns (SAPs). In this way, better error performance and a lower peak-to-average power ratio (PAPR) would be attained under proper system configurations, although in some cases, the achievable rate could be slightly reduced as compromise \cite{7469311,7509396,8353362}. Also, compared to spatial modulation (SM), OFDM-IM can get rid of the multi-antenna architecture and yields a low-cost and small-size implementation of IM \cite{7331321}.

Because OFDM-IM applies a unique modulation procedure to encode incoming bits, one of the most important aspects of OFDM-IM is the error performance, which can be characterized by average block error rate (BLER) and/or  bit error rate (BER). Therefore, they were investigated in most works related to OFDM-IM. However, in order to simplify the derivations and obtain closed-form expressions, the error performance analysis methodology in most existing works is based on two approximations stemming from the original work of the classic OFDM-IM presented in \cite{6587554}. The first one is to utilize the union bound assisted by the pairwise error probability (PEP) analysis. The second one is to employ an exponential approximation of Q-function.

Because of the employments of aforementioned two approximations, the accuracy of the analytical results of error performance for OFDM-IM loses and a gap exists between analytical and numerical results, especially for average BER. To enhance the analytical accuracy and explore the mechanism that causes the gap between analytical and numerical results, we abandon the exponential approximation of Q-function and adopt its exact expression in the form of Craig's formula \cite{258319}. By the Fubini–Tonelli theorem \cite{dibenedetto2002real}, the closed-form expressions of both average BLER and BER are obtained by this new error performance analysis methodology, and therefore easy to perform further analysis. As a result, the analytical accuracy will only lose due to the employed union bound with PEP analysis. Consequently, more accurate analytical results are achievable, and the impacts of both approximations on the gap between numerical and analytical results can be clarified.

\section{System Model}\label{sm}
The system model and assumptions adopted in this letter are exactly the same as those stipulated in \cite{6587554}. Meanwhile, to simplify the discussion and dedicate to the new error performance analysis methodology, we focus on only one single group consisting of $N$ subcarriers generated by inverse fast Fourier transform (IFFT) and interleaved grouping with independent and identically distributed (i.i.d.) faded channel power gains \cite{6841601}, among which $K$ subcarriers are activated to form a SAP, so as to modulate $\lfloor\log_2\binom{N}{K}\rfloor$-bit heading stream, where $\lfloor\cdot\rfloor$ denotes the floor function. To be simple, we follow the look-up table mapping relation specified in \cite{6587554} with lexicographical ordering. Specifically, $\binom{N}{K}$ SAPs are generated and listed in lexicographical order, and the first $2^{\lfloor\log_2\binom{N}{K}\rfloor}$ SAPs are adopted to construct a bijective mapping relation to the $2^{\lfloor\log_2\binom{N}{K}\rfloor}$ code words listed in lexicographical order. Furthermore, a sufficiently long cyclic prefix (CP) and perfect synchronization in both time and frequency domains are supposed in order to eliminate inter-symbol interference (ISI) and inter-channel interference (ICI). In this way, the OFDM-IM system can be considered and analyzed on a subcarrier-wise basis. For each active subcarrier, conventional amplitude-phase modulation (APM) is applied and we adopt $M$-ary phase-shift keying ($M$-PSK) in this letter, owing to its constant-envelope attribute. Consequently, $K\log_2(M)$-bit subsequent stream is modulated to be data constellation symbols conveyed on $K$ active subcarriers. Hence, $B=\lfloor\log_2\binom{N}{K}\rfloor+K\log_2(M)$ is the length of the entire bit stream and all these incoming bits are supposed to be equiprobable. For convenience, we denote the set of all $N$ subcarriers as $\mathcal{N}=\{1,2,\dots,N\}$ and set of $K$ active subcarriers as $\mathcal{K}(\sigma)$, where $\sigma$ is the index of the chosen OFDM block for transmission depending on the entire $B$-bit stream.

By IFFT and interleaved grouping, we can write the OFDM block for transmission as $\mathbf{x}(\sigma)=[x(\sigma,1),x(\sigma,2),\dots,x(\sigma,N)]^T\in\mathbb{C}^{N\times 1}$, where $x(\sigma,n)=\begin{cases}
\chi_n,~~~~\mathrm{if}~n\in\mathcal{K}(\sigma)\\
0,~~~~~~\mathrm{otherwise}\end{cases}$ and $\chi_n$ represents the $M$-ary data constellation symbol conveyed on the $n$th active subcarrier, which is normalized by $\chi_n\chi_n^*=1$ for simplicity. More details of the transmission module of OFDM-IM can be found in \cite{6587554}.

Propagating through i.i.d. Rayleigh fading channels over all subcarriers, the received OFDM block can be written as $\mathbf{y}(\sigma)=\sqrt{\frac{P_t}{K}}\mathbf{H}\mathbf{x}(\sigma)+\mathbf{w}\in\mathbb{C}^{N\times 1}$, where $P_t$ is the transmit power for the OFDM block, which is uniformly distributed over $K$ active subcarriers; $\mathbf{H}=\mathrm{diag}\{h(1),h(2),\dots,h(N)\}$ is the channel state matrix (CSM) consisting of the channel coefficients pertaining to $N$ subchannels $\{h(n)\}_{n=1}^{N}$ that characterize the channel state information (CSI), and the probability density function (PDF) and cumulative distribution function (CDF) with respect to the exponentially distributed channel power gain $G(n)=|h(n)|^2$ with mean $\mu$ are given by $f_g(\xi)=\frac{1}{\mu}\mathrm{exp}\left(-\frac{\xi}{\mu}\right)$ and $F_g(\xi)=1-\mathrm{exp}\left(-\frac{\xi}{\mu}\right)$ for Rayleigh fading channel model; $\mathbf{w}=[w(1),w(2),\dots,w(N)]^T$ is the vector of additive white Gaussian noise (AWGN) at the receiver, and $w(n)\sim\mathcal{CN}(0,N_0)$ is the random noise term on the $n$th subcarrier with the average noise power $N_0$.

In this letter, we apply the maximum-likelihood (ML) estimation at the OFDM receiver, and the estimated OFDM block is produced by
\begin{equation}\label{dsajhj2a223232}\small
\begin{split}
\hat{\mathbf{x}}(\hat{\sigma})=\underset{\dot{\mathbf{x}}(\dot{\sigma})\in\mathcal{X}}{\arg\min}\begin{Vmatrix}{\mathbf{y}}({\sigma})-\sqrt{\frac{P_t}{K}}\mathbf{H}\dot{\mathbf{x}}(\dot{\sigma})\end{Vmatrix}_F,
\end{split}
\end{equation}
where $\begin{Vmatrix}\cdot\end{Vmatrix}_F$ denotes the Frobenius norm of the enclosed matrix/vector and $\mathcal{X}$ is the set of all legitimate OFDM blocks for transmission by OFDM-IM. In this letter, we assume a slow fading scenario so that CSM $\mathbf{H}$ is supposed to be perfectly known at the receiver.

\section{Error Performance Analysis}
Because the information bits are modulated by both indices of active subcarriers and the data constellation symbols conveyed on $K$ active subcarriers, the entire OFDM block should be estimated holistically \cite{8241721}. Therefore, to characterize the error performance for OFDM-IM systems, one shall consider in a block level and/or in a bit level (only if the length of entire bit stream is invariant), which are characterized by average BLER and/or BER, respectively. Both measures are also highly associated and would be mutually transferable by a set of simple steps. In this section, we illustrate how to derive both in closed form by the proposed methodology based on Craig's formula and the Fubini–Tonelli theorem. 

\subsection{Definitions of Average BLER and BER}
\subsubsection{Average BLER}
Because the average BLER is more fundamental and average BER can be easily derived concomitantly \cite{8519769}, we first focus on the derivation of average BLER. In general, without involving any specific analytical methodology, we can express the conditional BLER on CSM $\mathbf{H}$ to be $P_e\left(\mathbf{x}(\sigma)\vert \mathbf{H}\right)=\mathbb{P}\left\lbrace\hat{\mathbf{x}}(\hat{\sigma})\neq \mathbf{x}(\sigma)\vert \mathbf{H}\right\rbrace$, where $\mathbb{P}\{\cdot\}$ denotes the probability of the random event enclosed. Following this, we attain the unconditional BLER by averaging $P_e\left(\mathbf{x}(\sigma)\vert \mathbf{H}\right)$ over CSM $\mathbf{H}$ and have $P_e\left(\mathbf{x}(\sigma)\right)=\underset{\mathbf{H}}{\mathbb{E}}\left\lbrace P_e\left(\mathbf{x}(\sigma)\vert \mathbf{H}\right)\right\rbrace$ for the OFDM block $\mathbf{x}(\sigma)$, where $\mathbb{E}\{\cdot\}$ denotes the expected value of the enclosed. To consider different OFDM blocks, we average it over $\mathbf{x}(\sigma)\in\mathcal{X}$, which yields the average BLER $\bar{P}_e=\underset{\mathbf{x}(\sigma)\in\mathcal{X}}{\mathbb{E}}\left\lbrace P_e\left(\mathbf{x}(\sigma)\right)\right\rbrace$.

\subsubsection{Average BER}
To characterize the error performance in a bit level, one can resort to the average BER. Likewise, the conditional BER on CSM $\mathbf{H}$ can be expressed by\footnote{This expression of conditional BER is only applicable for modulation schemes with fixed transmission rate $B$ in bit per channel use (bpcu).} $P_b\left(\mathbf{x}(\sigma)\vert \mathbf{H}\right)=\frac{1}{B}\mathbb{P}\left\lbrace\hat{\mathbf{x}}(\hat{\sigma})\neq \mathbf{x}(\sigma)\vert \mathbf{H}\right\rbrace\delta(\hat{\mathbf{x}}(\hat{\sigma})\neq \mathbf{x}(\sigma))$, where $\delta(\hat{\mathbf{x}}(\hat{\sigma})\neq \mathbf{x}(\sigma))$ represents the number of bit errors when the originally transmitted OFDM block $\mathbf{x}(\sigma)$ is erroneously estimated. Again, we ditto average it over CSM $\mathbf{H}$ and obtain $P_b\left(\mathbf{x}(\sigma)\right)=\underset{\mathbf{H}}{\mathbb{E}}\left\lbrace P_b\left(\mathbf{x}(\sigma)\vert \mathbf{H}\right)\right\rbrace$ for the OFDM block $\mathbf{x}(\sigma)$. Similarly, the average BER is produced by the following relation: $\bar{P}_b=\underset{\mathbf{x}(\sigma)\in\mathcal{X}}{\mathbb{E}}\left\lbrace P_b\left(\mathbf{x}(\sigma)\right)\right\rbrace$.

\subsection{Derivations of Average BLER and BER}
\subsubsection{Average BLER}
To derive the average BLER, we first dedicate to the derivation of its basic element, i.e., the conditional BLER on CSM $\mathbf{H}$ $P_e\left(\mathbf{x}(\sigma)\vert \mathbf{H}\right)$. We can employ the classic union bound approach assisted by PEP analysis and approximate $P_e\left(\mathbf{x}(\sigma)\vert \mathbf{H}\right)$ to be
\begin{equation}\label{dsakdjkasjdk223}\small
P_e\left(\mathbf{x}(\sigma)\vert \mathbf{H}\right)\approx\sum_{\hat{\mathbf{x}}(\hat{\sigma})\neq\mathbf{x}(\sigma)}P_e\left(\mathbf{x}(\sigma)\rightarrow \hat{\mathbf{x}}(\hat{\sigma})\vert\mathbf{H}\right),
\end{equation}
where $P_e\left(\mathbf{x}(\sigma)\rightarrow \hat{\mathbf{x}}(\hat{\sigma})\vert\mathbf{H}\right)$ represents the conditional PEP on CSM $\mathbf{H}$ that the originally transmitted OFDM block $\mathbf{x}(\sigma)$ is erroneously estimated to be $\hat{\mathbf{x}}(\hat{\sigma})$. By the ML estimation scheme specified in (\ref{dsajhj2a223232}), the conditional PEP $P_e\left(\mathbf{x}(\sigma)\rightarrow \hat{\mathbf{x}}(\hat{\sigma})\vert\mathbf{H}\right)$ can be written by Q-function $Q(x)=\frac{1}{2\pi}\int_{x}^{\infty}\mathrm{exp}\left(-u^2/2\right)\mathrm{d}u$ as
\begin{equation}\label{dsads5a452453232}\small
\begin{split}
P_e\left(\mathbf{x}(\sigma)\rightarrow \hat{\mathbf{x}}(\hat{\sigma})\vert\mathbf{H}\right)&=Q\left(\sqrt{\frac{P_t}{KN_0}\begin{Vmatrix}\mathbf{H}\left(\mathbf{x}(\sigma)-\hat{\mathbf{x}}(\hat{\sigma})\right)\end{Vmatrix}^2_F}\right)\\
&=Q\left(\sqrt{\frac{P_t}{KN_0}\sum_{n=1}^{N}G(n)\Delta(n,\sigma,\hat{\sigma})}\right),
\end{split}
\end{equation}
where $\Delta(n,\sigma,\hat{\sigma})=\left\lvert{x(\sigma,n)}-{x(\hat{\sigma},n)}\right\rvert^2$ for simplicity. 

By the original definition of Q-function, the argument is the lower limit of the interval of an integral, and thus it is demanding to carry out advanced analysis. To ease advanced analysis, an exponential approximation of Q-function $Q(x)\approx \frac{1}{12}\mathrm{exp}\left(-\frac{x^2}{2}\right)+\frac{1}{4}\mathrm{exp}\left(-\frac{2x^2}{3}\right)$ is adopted in the initial work of the classic OFDM-IM presented in \cite{6587554}, which gives a simple, neat and closed-form expression of the average error rate in both bit and block level. Therefore, most works analyzing error performance for OFDM-IM follow this approximation. However, because of this approximation as well as the union bound approach, the analytical accuracy reduces twice. Moreover, the simultaneous implementations of two approximations blur the mechanism causing the gap between analytical and numerical results. Consequently, we adopt Craig's formula $Q(x)=\frac{1}{\pi}\int_{0}^{\pi}\mathrm{exp}\left(-\frac{x^2}{2\sin^2\theta}\right)\mathrm{d}\theta$ to circumvent the approximation of Q-function \cite{258319}, and aim at obtaining a more accurate result as well as discovering the mechanism causing the gap between analytical and numerical results of the error performance for OFDM-IM. Accordingly, (\ref{dsads5a452453232}) can be rewritten as
\begin{equation}\label{craigseqqpro}\small
\begin{split}
&P_e\left(\mathbf{x}(\sigma)\rightarrow \hat{\mathbf{x}}(\hat{\sigma})\vert\mathbf{H}\right)=\frac{1}{\pi}\int_{0}^{\frac{\pi}{2}}\prod_{n=1}^{N}\mathrm{exp}\left(-\frac{P_tG(n)\Delta(n,\sigma,\hat{\sigma})}{2KN_0\sin^2\theta}\right)\mathrm{d}\theta.
\end{split}
\end{equation}
Following this, The unconditional PEP $P_e\left(\mathbf{x}(\sigma)\rightarrow \hat{\mathbf{x}}(\hat{\sigma})\right)$ can be determined in (\ref{zhiyouyigechangde}), where $(\mathrm{a})$ is valid according to the Fubini–Tonelli theorem\footnote{First, the integrand is a conditional probability and thereby non-negative. Second, the result of the multiple integral is a unconditional probability, which is bounded. Therefore, the application conditions of the Fubini–Tonelli theorem are satisfied.} that allows swapping order of integration \cite{dibenedetto2002real}; $(\mathrm{b})$ is derived by the independence among subcarriers by the system model assumed in this letter; $(\mathrm{c})$ is derived by partial fraction decomposition and the exchangeability between integral and summation operations; $\tau(n,\sigma,\hat{\sigma})=\frac{P_t\mu\Delta(n,\sigma,\hat{\sigma})}{2KN_0}$; $\{\alpha(n,\sigma,\hat{\sigma})\}_{n=1}^N$ is a set of constant coefficients depending only on $\{\tau(n,\sigma,\hat{\sigma})\}_{n=1}^N$ and is irrelevant to the coefficient $\theta$. More specifically, $\{\alpha(n,\sigma,\hat{\sigma})\}_{n=1}^N$ can be easily solved by the equation set infra:

\begin{figure*}[!t]
\begin{equation}\label{zhiyouyigechangde}\small
\begin{split}
&P_e\left(\mathbf{x}(\sigma)\rightarrow \hat{\mathbf{x}}(\hat{\sigma})\right)=\underset{\mathbf{H}}{\mathbb{E}}\left\lbrace P_e\left(\mathbf{x}(\sigma)\rightarrow \hat{\mathbf{x}}(\hat{\sigma})\vert\mathbf{H}\right)\right\rbrace=\frac{1}{\pi}\underset{\mathbf{H}}{\mathbb{E}}\left\lbrace\int_{0}^{\frac{\pi}{2}}\prod_{n=1}^{N}\mathrm{exp}\left(-\frac{P_tG(n)\Delta(n,\sigma,\hat{\sigma})}{2KN_0\sin^2\theta}\right)\mathrm{d}\theta\right\rbrace\\
&=\frac{1}{\pi}\underbrace{\int_{0}^{\infty}\int_{0}^{\infty}\dots\int_{0}^{\infty}}_{N-\mathrm{fold}}\left(\int_{0}^{\frac{\pi}{2}}\prod_{n=1}^{N}\mathrm{exp}\left(-\frac{P_tG(n)\Delta(n,\sigma,\hat{\sigma})}{2KN_0\sin^2\theta}\right)\mathrm{d}\theta\right)\left(\prod_{n=1}^{N}f_g(G(n))\right)\mathrm{d}G(1)\mathrm{d}G(2)\dots\mathrm{d}G(N)\\
&\overset{(\mathrm{a})}{=}\frac{1}{\pi}\int_{0}^{\frac{\pi}{2}}\left[\underbrace{\int_{0}^{\infty}\int_{0}^{\infty}\dots\int_{0}^{\infty}}_{N-\mathrm{fold}}\left(\prod_{n=1}^{N}\mathrm{exp}\left(-\frac{P_tG(n)\Delta(n,\sigma,\hat{\sigma})}{2KN_0\sin^2\theta}\right)\right)\left(\prod_{n=1}^{N}f_g(G(n))\right)\mathrm{d}G(1)\mathrm{d}G(2)\dots\mathrm{d}G(N)\right]\mathrm{d}\theta\\
&\overset{(\mathrm{b})}{=}\frac{1}{\pi}\int_{0}^{\frac{\pi}{2}}\left[\prod_{n=1}^{N}\left(\int_{0}^{\infty}\mathrm{exp}\left(-\frac{P_tG(n)\Delta(n,\sigma,\hat{\sigma})}{2KN_0\sin^2\theta}\right)f_g(G(n))\mathrm{d}G(n)\right)\right]\mathrm{d}\theta=\frac{1}{\pi}\int_{0}^{\frac{\pi}{2}}\left[\prod_{n=1}^{N}{(1+\tau(n,\sigma,\hat{\sigma})\csc^2\theta)^{-1}}\right]\mathrm{d}\theta\\
&\overset{(\mathrm{c})}{=}\frac{1}{\pi}\sum_{n=1}^{N}\left[\alpha(n,\sigma,\hat{\sigma})\int_{0}^{\frac{\pi}{2}}{(1+\tau(n,\sigma,\hat{\sigma})\csc^2\theta)^{-1}}\mathrm{d}\theta\right]=\frac{1}{2}\sum_{n=1}^{N}\left(\frac{\alpha(n,\sigma,\hat{\sigma})}{1+\tau(n,\sigma,\hat{\sigma})+\sqrt{\tau(n,\sigma,\hat{\sigma})(1+\tau(n,\sigma,\hat{\sigma}))}}\right)
\end{split}
\end{equation}
\hrule
\end{figure*}

\begin{equation}\small\label{dsakdjk2eusdsc}
\begin{cases}
\overset{N}{\underset{n=1}{\sum}}\alpha(n,\sigma,\hat{\sigma})=1\\
\left\lbrace\overset{N}{\underset{n=1}{\sum}}\left(\alpha(n,\sigma,\hat{\sigma})\underset{\nu\in \mathcal{V}({\mathcal{N}\setminus \{n\},i})}{\sum}\underset{u\in\nu}{\prod}\tau(u,\sigma,\hat{\sigma})\right)=0\right\rbrace_{i=1}^{N-1}
\end{cases}
\end{equation}
where $\mathcal{V}(\mathcal{S},i)$ denotes the set of all $i$-element subsets of a finite set $\mathcal{S}$. For example, if $\mathcal{S}=\{1,2,3,4\}$, then $\mathcal{V}(\mathcal{S},2)=\{\{1,2\},\{1,3\},\{1,4\},\{2,3\},\{2,4\},\{3,4\}\}$. Technically, the solution to (\ref{dsakdjk2eusdsc}) can be expressed in the form of matrix as $\mathbf{A}^{-1}\mathbf{b}$, where $\mathbf{A}$ is an $N\times N$ matrix whose elements in the first row is given by $a_{1j}=1$ and other elements are given by $a_{ij}=\underset{\nu\in \mathcal{V}({\mathcal{N}\setminus \{j\},i-1})}{\sum}\underset{u\in\nu}{\prod}\tau(u,\sigma,\hat{\sigma})$, $\forall~i\neq 1$; $\mathbf{b}=[1,0,0,\dots,0]^T$ is an $N\times 1$ vector. According to the rudiments of linear algebra and computational complexity theory, the computational complexity for solving (\ref{dsakdjk2eusdsc}) by matrix inversion is $\mathcal{O}(N^3)$, which indicates a polynomial time complexity and is acceptable for most practical applications.

For comparison purposes, we also approximate $P_e\left(\mathbf{x}(\sigma)\rightarrow \hat{\mathbf{x}}(\hat{\sigma})\right)$ by the conventional methodology using the exponential approximation of Q-function as follows:
\begin{equation}\small
P_e\left(\mathbf{x}(\sigma)\rightarrow \hat{\mathbf{x}}(\hat{\sigma})\right)\approx\sum_{i=1}^{2}\left[\rho_i\prod_{n=1}^{N}{(1+2\eta_i\tau(n,\sigma,\hat{\sigma}))^{-1}}\right],
\end{equation}
where $\{\rho_1,\rho_2\}=\{1/12,1/4\}$ and $\{\eta_1,\eta_2\}=\{1/2,2/3\}$.

Subsequently, by taking advantage of the additivity of expectation operation, we obtain the relation infra for the unconditional BLER $P_e\left(\mathbf{x}(\sigma)\right)$:
\begin{equation}\small
\begin{split}
&P_e\left(\mathbf{x}(\sigma)\right)\approx\underset{\mathbf{H}}{\mathbb{E}}\left\lbrace\sum_{\hat{\mathbf{x}}(\hat{\sigma})\neq\mathbf{x}(\sigma)}P_e\left(\mathbf{x}(\sigma)\rightarrow \hat{\mathbf{x}}(\hat{\sigma})\vert\mathbf{H}\right)\right\rbrace\\
&=\sum_{\hat{\mathbf{x}}(\hat{\sigma})\neq\mathbf{x}(\sigma)}\underset{\mathbf{H}}{\mathbb{E}}\left\lbrace P_e\left(\mathbf{x}(\sigma)\rightarrow \hat{\mathbf{x}}(\hat{\sigma})\vert\mathbf{H}\right)\right\rbrace=\sum_{\hat{\mathbf{x}}(\hat{\sigma})\neq\mathbf{x}(\sigma)}P_e\left(\mathbf{x}(\sigma)\rightarrow \hat{\mathbf{x}}(\hat{\sigma})\right).
\end{split}
\end{equation}
Finally, with the assumption that all incoming bits are equiprobable, we can average $P_e\left(\mathbf{x}(\sigma)\right)$ over all legitimate OFDM blocks $\mathbf{x}(\sigma)\in\mathcal{X}$ and obtain the average BLER as
\begin{equation}\small
\bar{P}_e=\frac{1}{X}\sum_{\mathbf{x}(\sigma)\in\mathcal{X}}P_e\left(\mathbf{x}(\sigma)\right),
\end{equation}
where $X=|\mathcal{X}|=2^{\lfloor\log_2\binom{N}{K}\rfloor}M^K$ is the number of all legitimate OFDM blocks.

\subsubsection{Average BER}
Similarly for average BER, we can approximate it simply by the union bound approach based on the unconditional PEP derived in (\ref{zhiyouyigechangde}), and obtain
\begin{equation}\small
\bar{P}_b=\frac{1}{XB}\sum_{\mathbf{x}(\sigma)\in\mathcal{X}}\sum_{\hat{\mathbf{x}}(\hat{\sigma})\neq\mathbf{x}(\sigma)}P_e\left(\mathbf{x}(\sigma)\rightarrow \hat{\mathbf{x}}(\hat{\sigma})\right)\delta\left(\mathbf{x}(\sigma)\rightarrow \hat{\mathbf{x}}(\hat{\sigma})\right),
\end{equation}
where $\delta\left(\mathbf{x}(\sigma)\rightarrow \hat{\mathbf{x}}(\hat{\sigma})\right)$ represents the number of bit errors for a single pairwise error event that $\mathbf{x}(\sigma)$ is erroneously estimated to be $\hat{\mathbf{x}}(\hat{\sigma})$.

\section{Numerical Results and Discussion}
\begin{figure}[!t]
\centering
\includegraphics[width=5.2in]{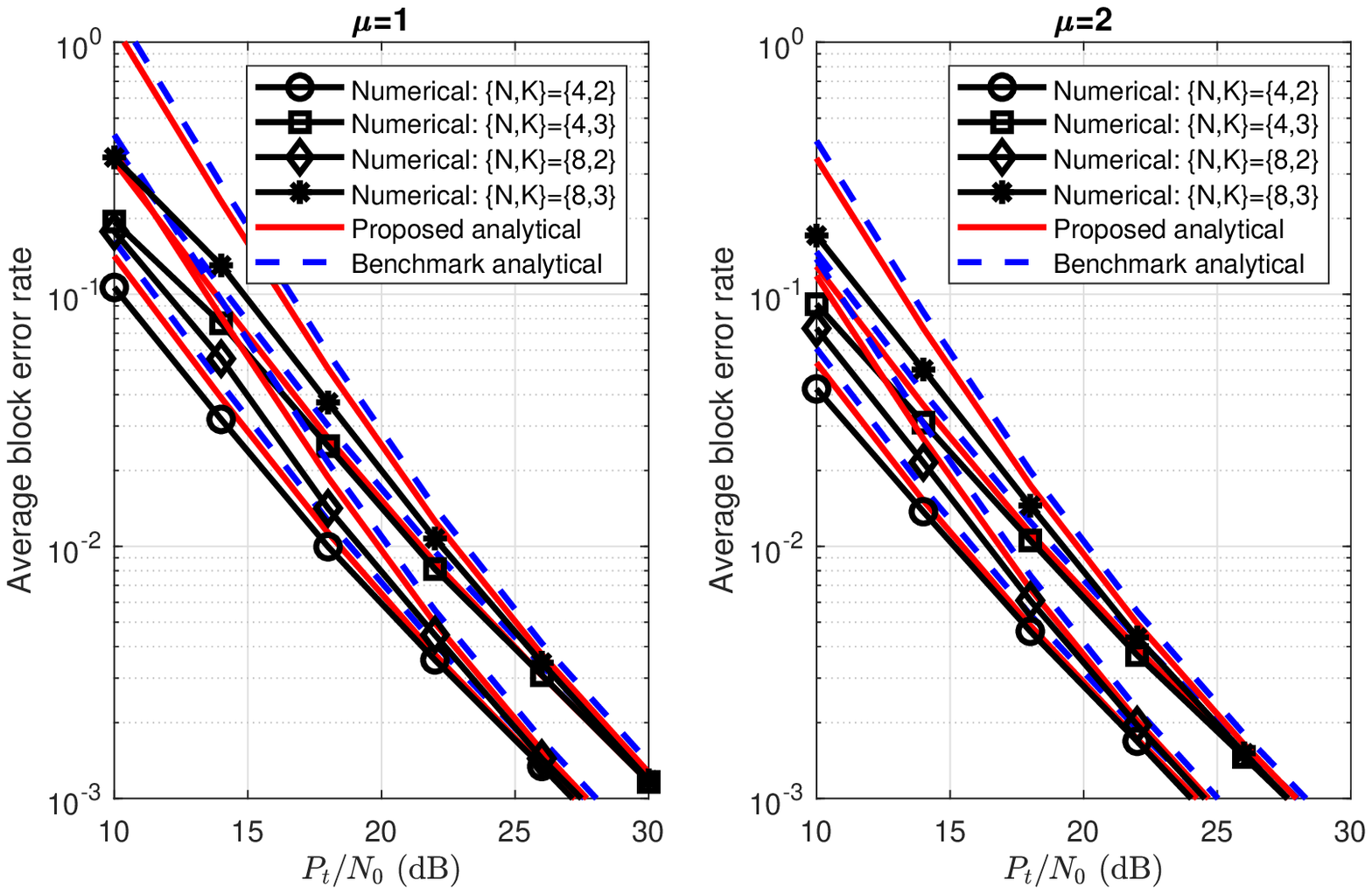}
\caption{Average BLER vs. ratio of transmit power to noise power $P_t/N_0$.}
\label{BLER}
\end{figure}

\begin{figure}[!t]
\centering
\includegraphics[width=5.2in]{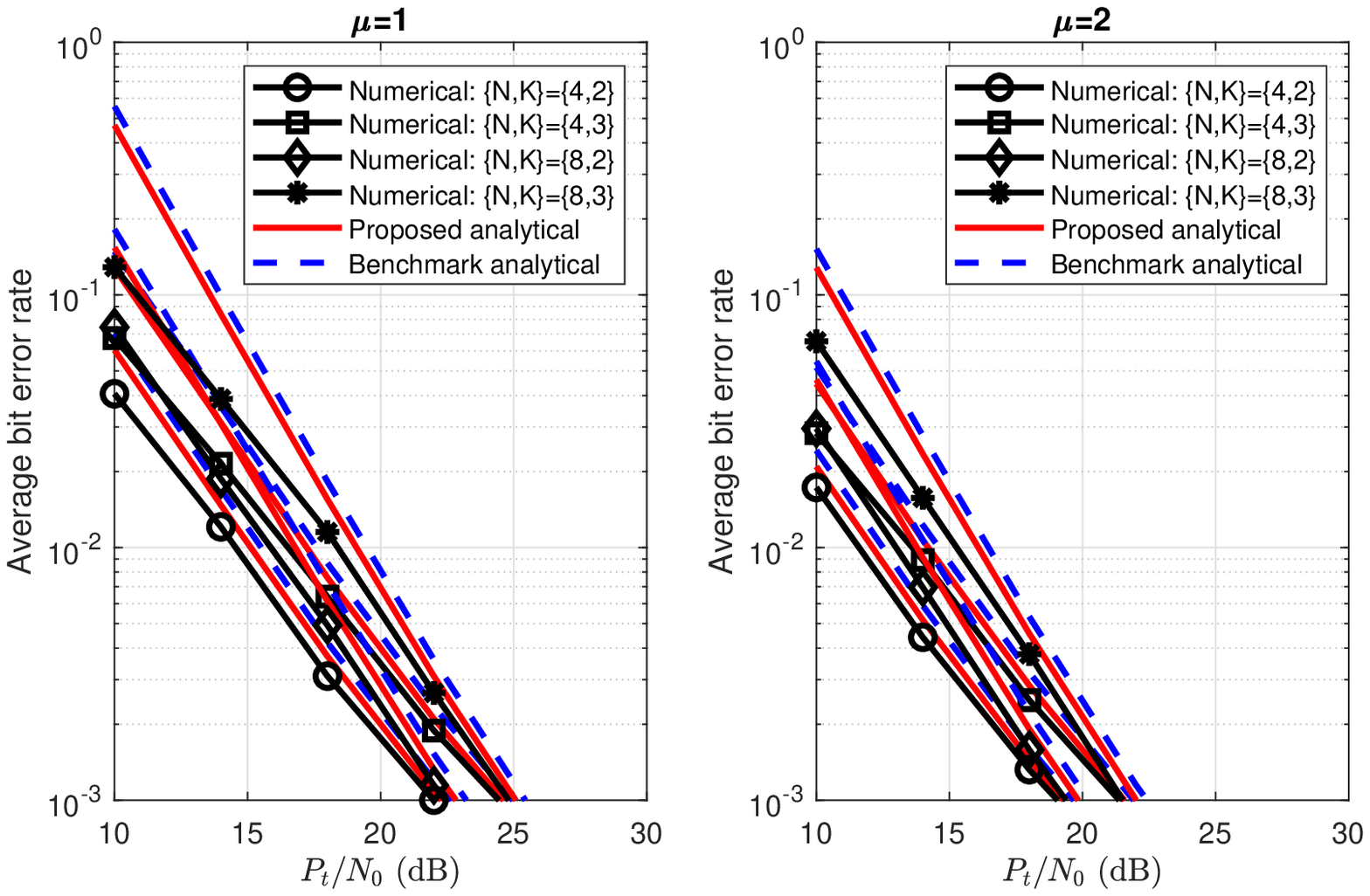}
\caption{Average BER vs. ratio of transmit power to noise power $P_t/N_0$.}
\label{BER}
\end{figure}

To verify the proposed error performance analysis methodology based on Craig's formula and compare it with the conventional methodology relying on the exponential approximation of $Q(x)$, we carried out a set of simulations by Monte Carlo methods. Simulation results are presented and discussed in this section. For simplicity, we set the fading channel with average channel power gain $\mu=1,2$ and adopt binary PSK (BPSK) as the APM scheme as well as Gray code for bit mapping. We take the analytical results generated by the conventional methodology relying on the exponential approximation of Q-function as comparison benchmarks in our simulations. Simulation results pertaining to average BLER and BER are illustrated in Fig. \ref{BLER} and Fig. \ref{BER}, respectively. From both figures, we can observe that the analytical accuracy has been improved for both average BLER and BER, as the gaps between analytical and numerical results narrow in comparison with those provided by the conventional methodology. Besides, it can also be found that the union bound assisted by PEP analysis yields a significant impact on error performance when $P_t/N_0$ is small.  Except $P_t/N_0$, other system parameters have only a negligible impact on the accuracy of the proposed method based on Craig's formula.

Note that, we only take the classic OFDM-IM as an example to illustrate the feasibility of the proposed methodology based on Craig's formula, although the improvement on accuracy is not significant in this simplistic case. However, as shown in some literature considering advanced OFDM-IM transmission scenarios, the error caused by the approximation of Q-function would be magnified, especially when channel ordering and multi-hop transmission are involved \cite{8241721,8519769}. More important, another key attribute of the proposed methodology is its convergence. The approximate results based on the exponential approximation of Q-function will always yield a fixed gap to the numerical results in the high signal-to-noise ratio (SNR) region, which does not diminish when increasing $P_t/N_0$. On the contrary, the performance gap becomes smaller by the Craig's formula based methodology when increasing $P_t/N_0$, owing to the convergence of the union bound at high SNR.

\section{Conclusion}\label{c}
In order to obtain more accurate analytical results pertaining to the error performance for OFDM-IM, we proposed a new error performance analysis methodology based on Craig's formula for OFDM-IM. We derived the average BLER and BER in closed form by the proposed analytical methodology and examined them by numerical simulations. Moreover, the impacts of both approximations, i.e., the union bound and the exponential approximation of Q-function utilized in most existing works for error performance analysis can be clarified.

\bibliographystyle{IEEEtran}
\bibliography{bib}

\end{document}